\documentclass{PoS}
 
\def\lsi{\raise0.3ex\hbox{$<$\kern-0.75em\raise-1.1ex\hbox{$\sim$}}}
\def\gsi{\raise0.3ex\hbox{$>$\kern-0.75em\raise-1.1ex\hbox{$\sim$}}}
\newcommand{\lsim}{\mathop{\lsi}}
\newcommand{\gsim}{\mathop{\gsi}}
\newcommand{\tr}{\mathop{\rm Tr}}
\newcommand{\Hc}{{\rm H.c.\ }}
\newcommand{\Tc}{T_{\rm c}}
\newcommand{\msbar}{{\overline{\mbox{\rm MS}}}}
\newcommand{\fig}{fig.~}

\title{First order thermal phase transition with 126 GeV Higgs mass}

\ShortTitle{First order thermal phase transition with 126 GeV Higgs mass}

\author{M.~Laine$^a$,
        G.~Nardini$^{b,c}$,  
        \speaker{K.~Rummukainen}${\,}^{d,e}$\\
        \llap{$^a$}
        Institute for Theoretical Physics, AEC, 
        University of Bern, CH-3012 Bern, Switzerland\\
        \llap{$^b$}
        Faculty of Physics, University of Bielefeld, 
        D-33501 Bielefeld, Germany\\
        \llap{$^c$}
        DESY, Notkestr.\ 85, D-22607 Hamburg, Germany\\
        \llap{$^d$}
        Department of Physics,
        P.O.Box 64, FI-00014 University of Helsinki, Finland\\
        \llap{$^e$}
        Helsinki Institute of Physics,
        P.O.Box 64, FI-00014 University of Helsinki, Finland\\
        E-mail: 
        \email{laine@itp.unibe.ch}, 
        \email{germano.nardini@desy.de}, 
        \email{kari.rummukainen@helsinki.fi}
        }

      \abstract{We study the strength of the electroweak phase
        transition in models with two light Higgs doublets and a light
        SU(3)$_c$ triplet by means of lattice simulations in 
        a dimensionally reduced effective theory. In the parameter region
        considered the transition on the lattice is significantly
        stronger than indicated by a 2-loop perturbative
        analysis. Within some ultraviolet uncertainties, the finding
        applies to MSSM with a Higgs mass $m_h \approx 126$\,GeV and
        shows that the parameter region useful for electroweak baryogenesis
        is enlarged. In particular (even though
        only dedicated analyses can quantify the issue), the tension
        between LHC constraints after the 7~TeV and 8~TeV runs
        and frameworks where the electroweak phase transition 
        is driven by light stops, seems to be relaxed.}

\FullConference{31st International Symposium on Lattice Field Theory - LATTICE 2013\\
		July 29 - August 3, 2013\\
		Mainz, Germany}

\begin{document}

%
\section{Introduction}

Electroweak baryogenesis (EWBG) \cite{krs} is an attractive scenario for
the generation of the baryon number of the Universe.  It requires the
existence of a relatively strong first order thermal electroweak phase
transition.  As the Universe cools down from temperatures above the
electroweak scale $T\approx T_{\rm EW} \sim 100$\,GeV, the
metastability associated with first order transitions leads to thermal
non-equilibrium, which is one of the necessary Sakharov conditions for
successful baryogenesis.  The other conditions (C, CP and baryon
number violation) exist in the Standard Model (SM) and its simple
extensions.  Therefore, it is well motivated to study the
thermodynamics of the electroweak ``symmetry breaking''
phase transition in these theories.

It was established already more than 15 years ago, 
however, that actually the
SM does not have a strong first order phase transition.
Indeed, in a series of lattice simulations
\cite{endpoint}--\cite{buwu} 
it was unambiguously shown that the transition is a smooth cross-over at
Higgs masses larger than about 
72\,GeV.  This value was well below the LEP limits of the time.  Thus,
it was necessary to look beyond the SM for possibilities for EWBG.

A strong first order phase transition is possible in several
extensions of the SM, in particular in the Minimal Supersymmetric
Standard Model (MSSM) if the right-handed stop, the scalar partner of
the top quark, is sufficiently light and the left-handed stop is heavy
\cite{Carena:1996wj}--\cite{mlo2}.
Although tightly constrained, perturbative analyses of phase
transition properties as well as of current LHC bounds indicate that
the parameter space for MSSM EWBG may still be open
\cite{Carena:2008vj}--\cite{new3}.

However, the accuracy of perturbative analyses of phase transition
properties is limited by the infrared singularities inherent to
thermal field theory \cite{linde}: the physics of the momentum scales
$p~\!\!\sim~\!\!g^2 T / \pi $ is non-perturbative, even at weak coupling $g
\ll \pi$.  For reliable results it is thus necessary to use numerical
lattice simulations.  A striking example of these problems is provided
by the phase transition in the SM: perturbation theory predicts that
the transition becomes weaker but remains of first order as the Higgs
mass increases, in contrast to the lattice results mentioned above
which indicate that the transition ceases to exist for $m_H >
72$\,GeV.

Three-dimensional (3d) dimensionally reduced {\em effective theories}
provide for a particularly convenient and successful tool for studying
the thermodynamics of weakly coupled theories on the lattice.  The
effective theory is derived from the original four-dimensional (4d)
theory using perturbative methods, in a computation which suffers from
no infrared problems. The infrared sector of the original theory is
fully transferred to the effective theory where it can be studied by
lattice simulations (just in three dimensions and thus less demanding
than in four, particularly recalling that both light and heavy chiral
fermions are present in the SM and its extensions).  Most of the above
results concerning the SM phase transition were obtained using the
effective theory approach \cite{endpoint}--\cite{Gurtler:1997hr}.
It has also been
successfully applied to high-temperature QCD
(cf.\ e.g. refs.~\cite{Kajantie:1998yc,Hietanen:2008tv}), although in that case
the accuracy is limited by the larger gauge coupling.

The effective theory approach was used to study the phase transition
in the MSSM in a series of papers in 1998--2001
\cite{mssmsim}--\cite{cpsim}.  
The parameters used in those
computations were chosen following the ever increasing LEP bounds, 
eventually extending up to 
$m_H \sim 115$\,GeV.
The experimental discovery of the Higgs particle at
$126$\,GeV thus motivates us to revisit the MSSM phase transition on
the lattice. This was recently 
achieved in ref.~\cite{own}, whose results we review here.

The results obtained should, however, be more generic than just for
the specific case of MSSM. Indeed, we expect that {\it (i)}
the electroweak phase transition is stronger than perturbatively
estimated in a broad neighbourhood of the effective-theory parameter
point analyzed here, and that {\it (ii)} several 4d theories with phase
transitions driven by light colored (SU(3)$_c$ triplet) scalars can
fall into this neighbourhood. Trivial examples are non-minimal
supersymmetric models with a light scalar sector consisting of two Higgs
SU(2)$_L$ doublets and a right-handed stop.

%
\section{Effective theory}

Dimensionally reduced effective theories are by now well established
as reliable tools to study weakly coupled field theories at high
temperatures.  Starting from the original 4d theory at finite
temperature, the momentum modes $p \sim \pi T$ and $p \sim gT$ are
integrated out in stages.  The resulting theory contains only the
non-perturbative soft sector $p\sim g^2 T / \pi$.  Because all
non-zero Matsubara modes have $p \sim \pi T$, all fermion modes are
integrated out and the resulting theory lives in 3d (dimensional
reduction).  The parameters of the effective theory depend on the
parameters of the 4d theory, which in turn are set by physical
observables (e.g. pole masses).

We consider a 4d theory with weak SU(2)$_L$ and strong SU(3)$_c$ gauge
fields, two colorless Higgs SU(2)$_L$ doublet scalars, $H_1$ and
$H_2$, and an SU(3)$_c$ triplet scalar which is SU(2)$_L$ singlet. The
masses of the scalars are at the electroweak scale.  The 3d Lagrangian
describing this theory has the most general form allowed by
symmetries:
\begin{eqnarray*}
\label{effLagr}
{\cal L}_{\rm 3d} & = & 
 \frac12 \tr G_{ij}^2 + (D_i^s U)^\dagger (D_i^s U) + 
 m_U^2\, U^\dagger U  
 + \lambda^{ }_U\, (U^\dagger U )^2 \\
 & + & 
 \gamma^{ }_1\, U^\dagger U H_1^\dagger H^{ }_1 + 
 \gamma^{ }_2\, U^\dagger U H_2^\dagger H^{ }_2
 + \Bigl[\gamma_{12}\, U^\dagger U H_1^\dagger  H^{ }_2+\Hc\Bigr] \\
& + & 
 \frac12 \tr F_{ij}^2 
 + (D_i^w H^{ }_1)^\dagger(D_i^w H^{ }_1)
 + (D_i^w H^{ }_2)^\dagger(D_i^w H^{ }_2) \\
& + &  m_1^2\, H_1^\dagger H^{ }_1 +  m_2^2\, H_2^\dagger H^{ }_2 
 + \Bigl[ m_{12}^2\, H_1^\dagger  H^{ }_2 +\Hc  \Bigr]  \\
& + &  \lambda_1\, (H_1^\dagger H^{ }_1)^2 + 
 \lambda_2\, (H_2^\dagger H^{ }_2)^2 + 
 \lambda_3\, H_1^\dagger H^{ }_1 H_2^\dagger H^{ }_2 + 
 \lambda_4\, H_1^\dagger  H^{ }_2  H_2^\dagger H^{} _1 \\
& + & \Bigl[ \lambda_5\, (H_1^\dagger  H^{ }_2)^2 + 
 \lambda_6\, H_1^\dagger H^{ }_1 H_1^\dagger H^{ }_2 +
 \lambda_7\, H_2^\dagger H^{ }_2 H_1^\dagger H^{ }_2 + \Hc\Bigr]
 \;. 
\end{eqnarray*}
Here $G_{ij}$ and $F_{ij}$ are the SU(3)$_c$ and SU(2)$_L$ field strength
tensors, respectively, and the parameters $g_w^2,\, g_s^2,\, m_i^2, \,
\lambda_j$ and $\gamma_i$ have a
well-defined dependence on the original 4d parameters (including the
temperature $T$).  

Had we added further particles in the 4d theory that are Boltzmann
suppressed during the transition, or extra light fermions that do not 
become strongly coupled in either of the phases, 
then the 3d Lagrangian would
still be the same and only the relations between 3d and 4d parameters
would change. An example of a theory leading to the above
${\cal L}_{\rm 3d}$
is the MSSM when both Higgses and
right-handed stops are light (to a good approximation, further MSSM
scalars could be light too if weakly coupled to the light stop-Higgs
sector). In such a case the 3d and 4d parameters are related as
described in refs.~\cite{own, cpdr}. We focus on this model
in the following.

We fix the parameters so that the masses of the lightest CP-even Higgs
and right-handed stop are $m_H \approx 126$\,GeV and $m_{\tilde t_R}
\approx 155$\,GeV, respectively. Other squarks are heavy, $m_Q \gsim
7$\,TeV. For the remaining parameters we refer to ref.~\cite{own}.
The large scale $m_Q$ may induce large logarithms that have to be
resummed for precise relations between 4d parameters and physical
observables~\cite{Carena:2008rt}. We do not perform this
refinement here and uncertainties of several GeV may be present,
especially in $m_{\tilde t_R}$. This ambiguity could be reduced with
full 2-loop dimensional reduction, as was done for the SM
\cite{generic}. To be reminded of the uncertainties we label 
the 4d parameters with an asterisk (*).

It should be stressed, however, that 
the ambiguity affects only 4d parameters.
The theory we analyze by 3d perturbation
theory and by 3d lattice simulations, although it might correspond to
slightly different 4d observables than anticipated, 
is exactly the same. The
comparison of perturbation theory and lattice results is therefore
unambiguous within the effective theory.

%
\section{Simulations and results}

The lattice discretization of the theory is described in
ref.~\cite{cpsim}.  Because the effective theory is super-renormalizable
and all counterterms are known, the lattice parameters
do not require any tuning and in the continuum limit the
results can be directly compared with perturbative ones.

%
\begin{table}
\centerline{\begin{tabular}{cl}
\hline
${\beta_w = 4/(g_w^2 T^* a)}$ & volumes \\
\hline
~8     & $12^3$, $16^3$ \\
10     & $16^3$  \\
12     & $16^3$, $20^3$, $32^3$, $12^2\times 36$, $20^2\times 40$ \\
14     & $24^3$, $14^2\times 42$, $24^2\times 48$ \\
16     & $24^3$, $16^2\times 48$, $20^2\times60$, $24^2\times 72$ \\
20     & $32^3$, $20^2\times 60$, $26^2\times72$, $32^2\times64$ \\
24     & $24^3$, $32^3$, $48^3$, $24^2\times78$, $30^2\times72$ \\
30     & $48^3$ \\
\hline
\end{tabular}}
\caption[a]{Lattice spacings and volumes used in the simulations
of ref.~\cite{own}.}
\label{volumes}
\end{table}
%

The lattice spacing $a$ is parameterized through the SU(2)$_L$ 
gauge coupling: $\beta_w = 4/(g_w^2 T^* a)$.
The simulation volumes are listed in table \ref{volumes}.
Because the theory is fully bosonic, the simulations are inexpensive
and we can access a range of almost four in lattice spacings.
The update algorithm is a combination of heat bath and over-relaxation
updates.

%
\begin{figure}[t]


\centerline{%
\includegraphics[width=0.42\textwidth]{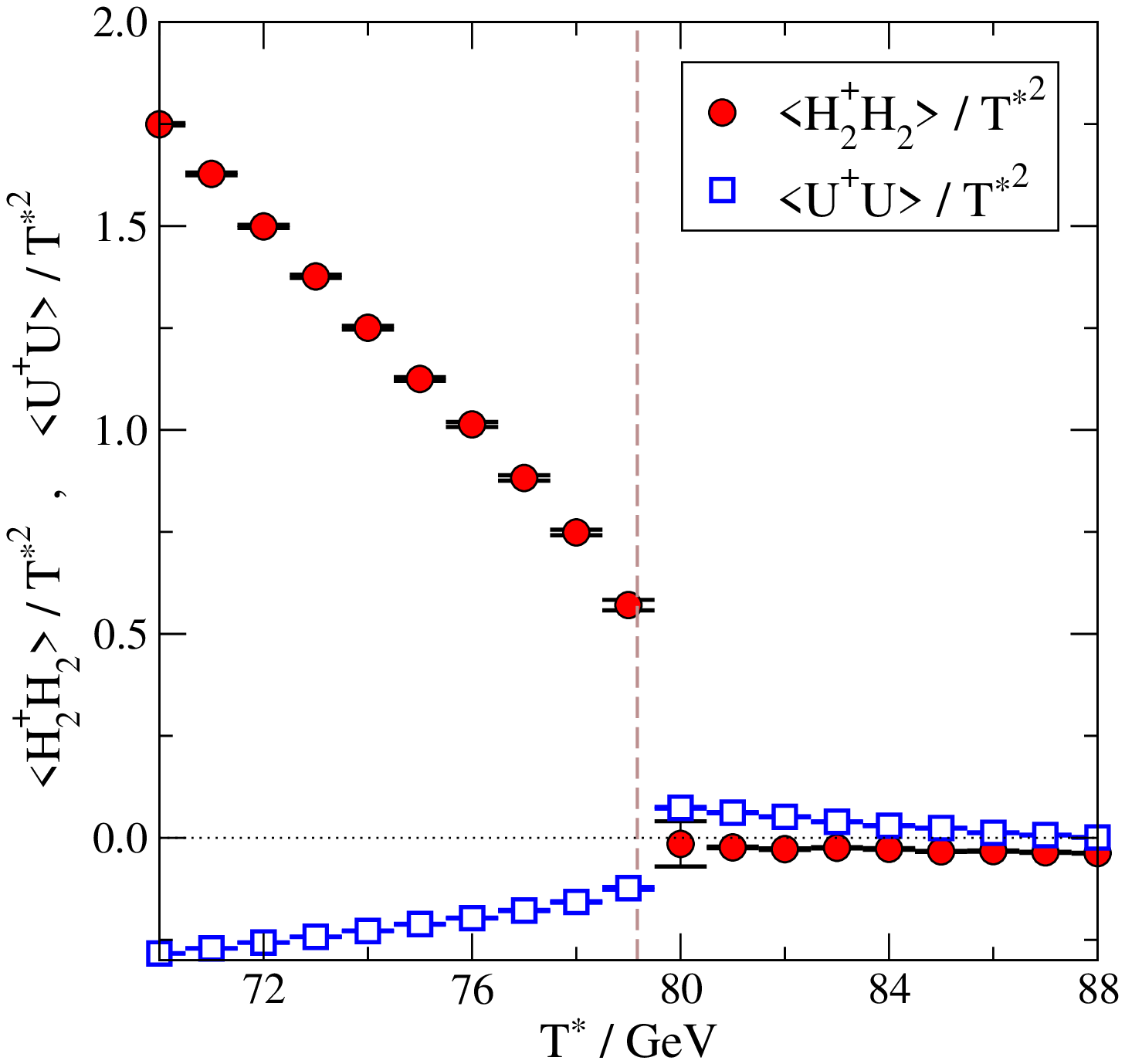}%
\qquad\includegraphics[width=0.42\textwidth]{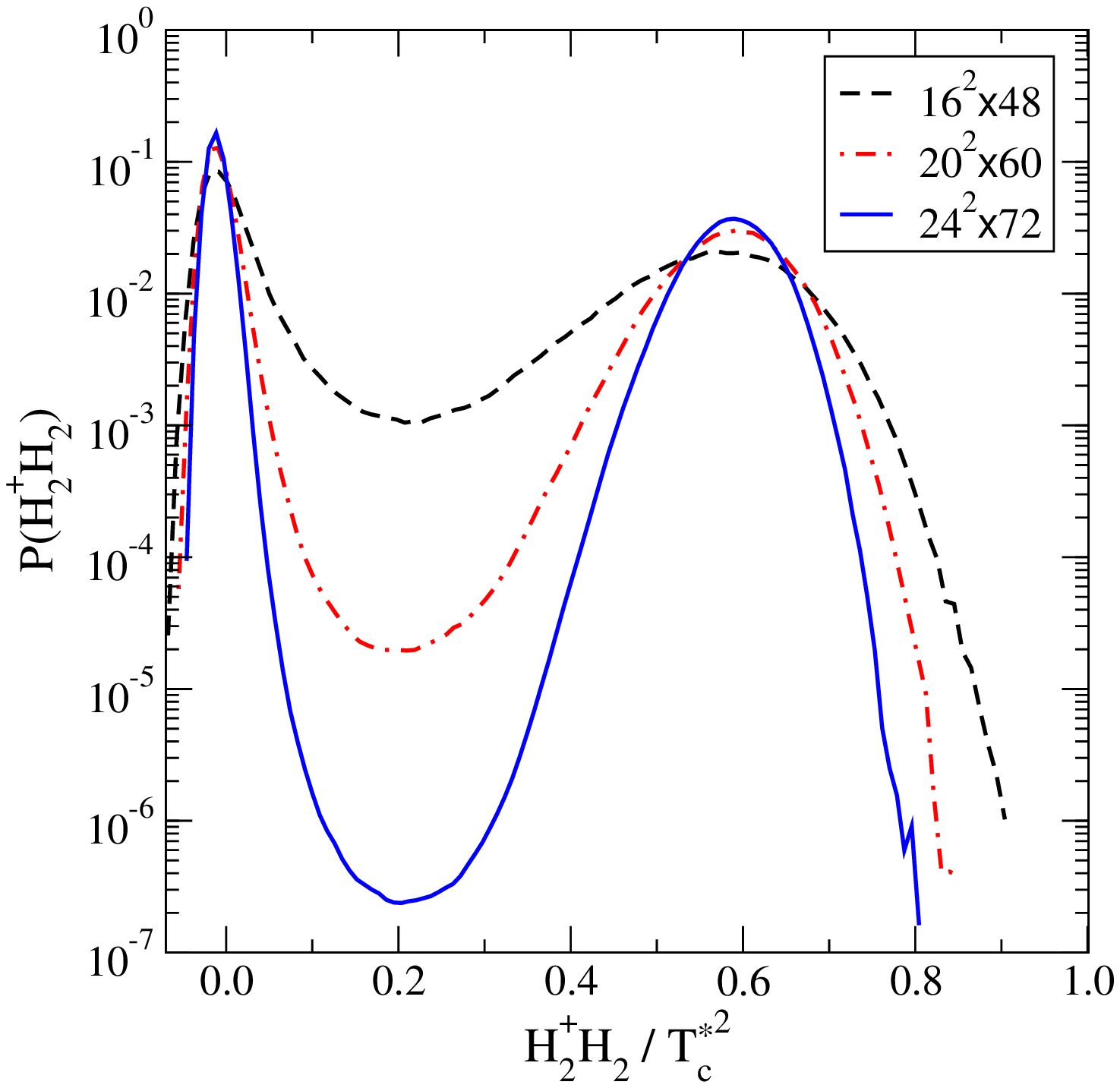}%
}
\caption[a]{Left: continuum-extrapolated expectation values of 
$\langle H_2^\dagger H^{ }_2\rangle$ 
and  
$\langle U^\dagger U\rangle$, both renormalized
in the $\msbar$ scheme, 
as functions of the temperature.
Right: the probability distribution of 
$\langle H_2^\dagger H^{ }_2\rangle$ at $\Tc^*$ using
different volumes. Both plots are from ref.~\cite{own}.}
\label{fig:condensates}
\end{figure}
%

In \fig\ref{fig:condensates}(left) we show a temperature scan of 
the continuum-extrapolated  $H_2$ and stop condensates. ($H_1$ is
much heavier and in practice inert in the transition.)  The
first-order nature is evident, as is the fact that the stop
field responds to the jump in the Higgs condensate.

\begin{figure}[t]


\centerline{%
\includegraphics[width=0.42\textwidth]{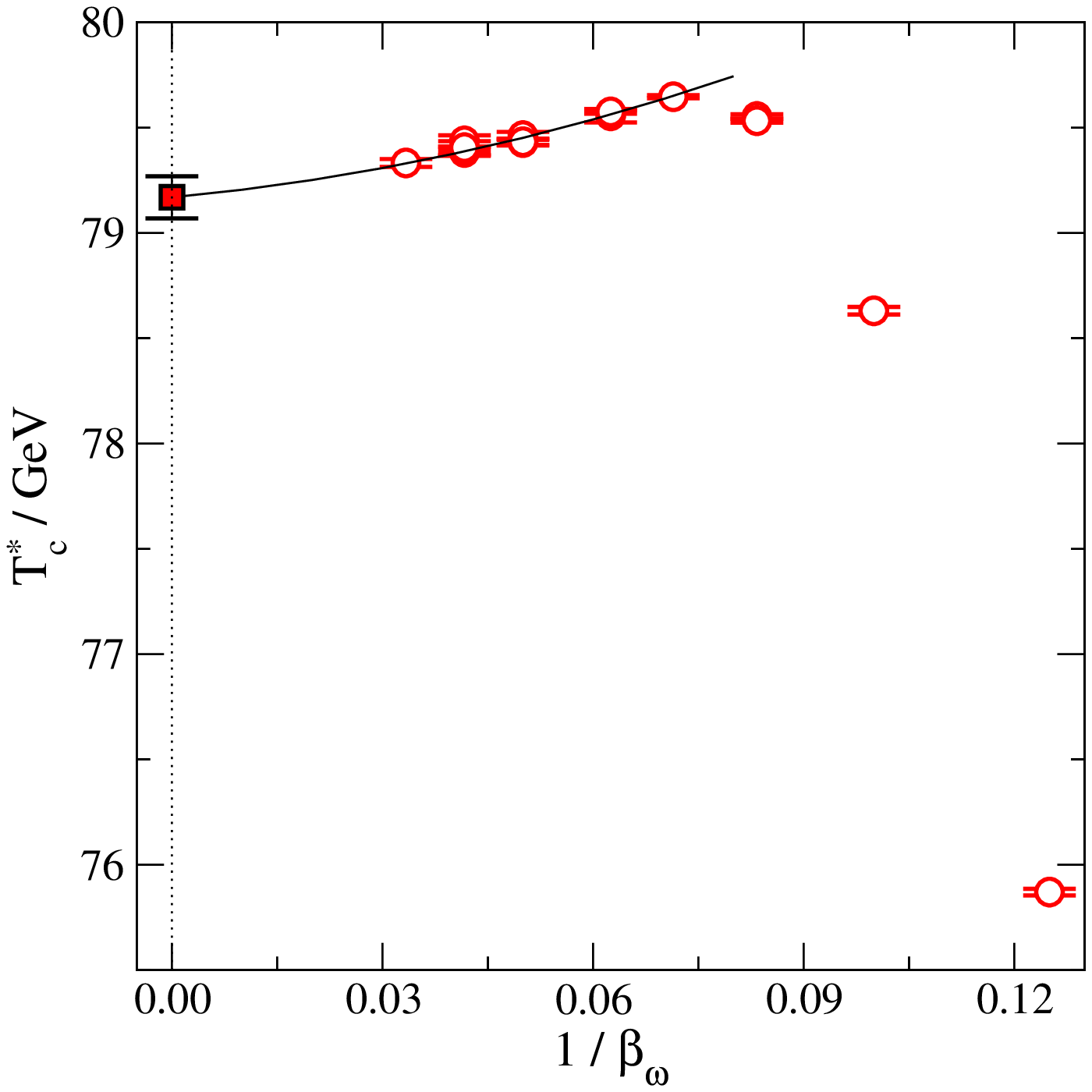}%
\qquad\includegraphics[width=0.42\textwidth]{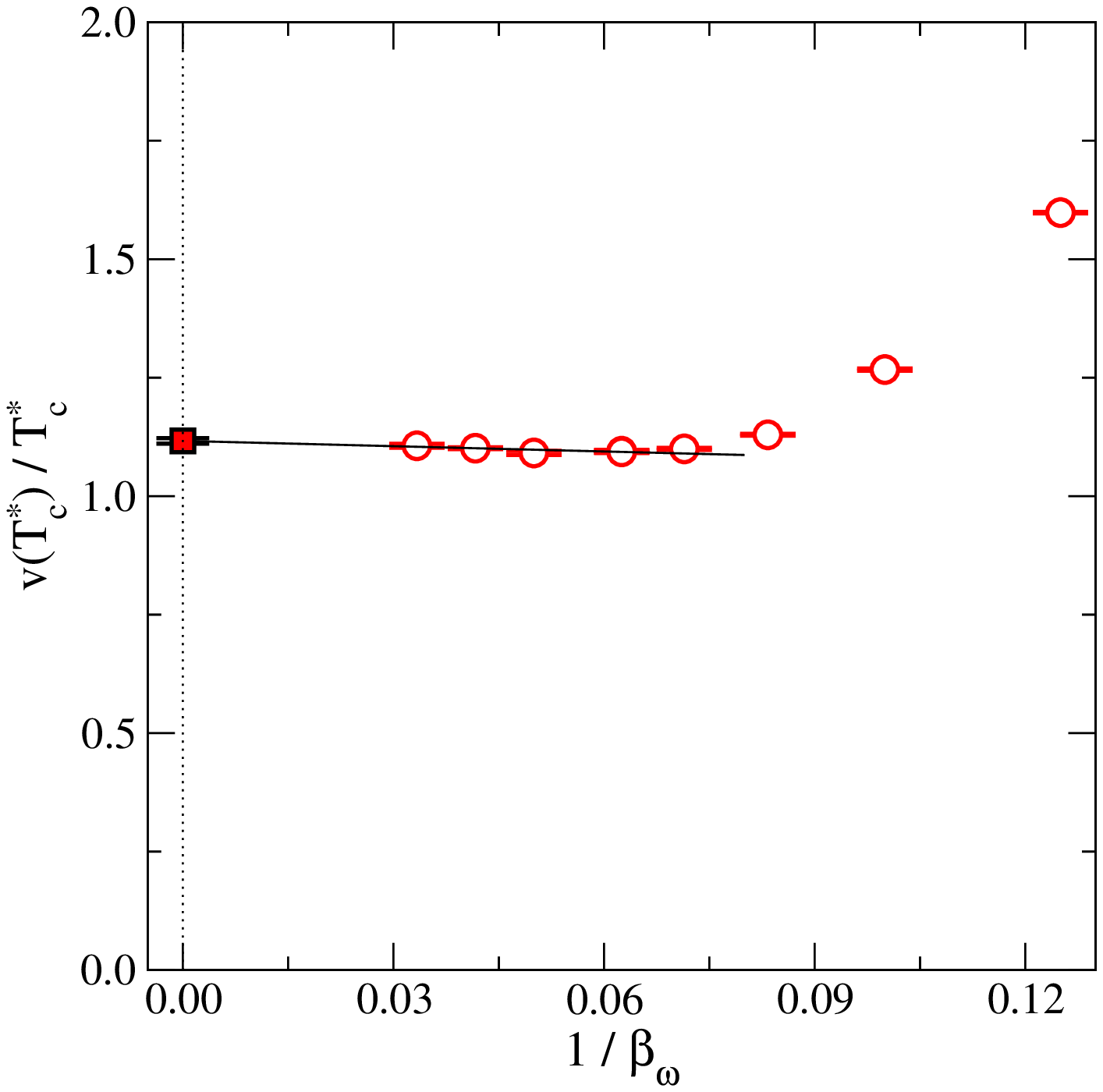}%
}
\caption[a]{Continuum extrapolations for the critical temperature (left)
  and the Higgs condensate (right), the latter defined on the lattice
as $v^2 \equiv 2 \Delta \langle \sum_{i=1}^{2} H_i^\dagger H^{ }_i \rangle$.  
Both plots are from ref.~\cite{own}.}
\label{fig:extrap}
\end{figure}

Most of our simulations are performed at the critical temperature,
employing multicanonical techniques as described in ref.~\cite{cpsim}.
Some of the resulting probability distributions 
of $\langle H_2^\dagger H^{ }_2\rangle$
are shown in \fig\ref{fig:condensates}(right).  These distributions
enable precise measurements of the critical temperature (equal
area of the peaks), Higgs condensate discontinuity $v(\Tc^*)$, the
latent heat of the transition and the tension of the interface
between the symmetric and broken phases.  The continuum extrapolations
of the critical temperature and the Higgs condensate are
shown in \fig\ref{fig:extrap}, using a linear + quadratic fit.
At $\beta_w = 4/(g_w^2 a T^*) \ge 14$ the cutoff effects are small
and a robust continuum limit can be obtained.  At coarser lattice
spacings the cutoff effects are sizable, which can be attributed
to the rather heavy mass of the $H_1$ field.

%
\begin{table}[t]
\centerline{
\begin{tabular}{lll}
     &   lattice & 2-loop \\
\hline
Transition temperature  {$\Tc^*$/GeV} & 79.17(10)  &  84.4   \\
Higgs discontinuity   {$v/\Tc^*$}   & 1.117(5)   &  0.9    \\
Latent heat          {$L/(\Tc^*)^4$}   & 0.443(4)   &  0.26   \\
Surface tension      {$\sigma/(\Tc^*)^3$} & 0.035(5) & 0.025  \\
\hline
\end{tabular}}
\caption[a]{Comparison of lattice and perturbative results, 
 both for the same 3d effective parametrization~\cite{own}.}
\label{table:comparison}
\end{table}
%

The numerical results are compared with 2-loop perturbative
ones in table~\ref{table:comparison}.
The transition is significantly stronger than indicated
by the perturbative analysis.  This is also evidenced by the Higgs
condensate discontinuity shown in \fig\ref{fig:compare}:
the transition becomes stronger as we go from 1 to 2 loops and then to 
lattice, and the transition temperature is simultaneously decreased.
This is in line with the previous experiences at $m_H \lsim 115$\,GeV
\cite{cpsim}. 

%
%

%
\section{Conclusions}

By means of dimensional reduction and 3d lattice simulations, we have
studied the finite temperature phase transition in models with two
Higgs SU(2)$_L$ doublets and a SU(3)$_c$ triplet as light scalar
degrees of freedom. In the parameter region considered, the transition
on the lattice is stronger than indicated by 2-loop perturbation
theory. In particular, it is strong enough for EWBG: as confirmed by
real-time simulations within the SM~\cite{db}--\cite{mdo} the
sphaleron rate is strongly correlated with $v/T$ and for the value
quoted in table~\ref{table:comparison} so suppressed that (if no
sizeable magnetic background is
present~\cite{Comelli:1999gt,DeSimone:2011ek}) it does not erase any
baryon asymmetry generated. A strong transition also implies
significant supercooling and non-trivial dynamics, leading to a
gravitational wave signal~\cite{Hindmarsh:2013xza}, although probably
not large enough to be observable in the foreseeable future.

Our result applies to any model with the above light scalar fields
participating in the transition, but in particular it 
covers many supersymmetric
frameworks. Within some ultraviolet uncertainties, it
shows that there seems to be room for EWBG in MSSM with $m_H\approx
126\,$GeV and $m_{\tilde t_R}\lesssim 155\,$GeV. If the latter bound
were confirmed by more precise analyses, MSSM baryogenesis would be 
less in tension with LHC data than previously estimated, with
$m_{\tilde t_R}\lesssim 110\,$GeV~\cite{cnqw}. Indeed, constraints
from Higgs searches would relax as stop enhancement in Higgs gluon
fusion would reduce, and loopholes in LHC stop analyses \cite{cnqw}
might open up.


%
\vspace{2mm}
{\em Acknowledgement:}
K.R acknowledges support from the Academy of Finland project
1134018.

%

\end{document}